\documentclass[aps,prb,twocolumn,showpacs,groupedaddress,preprintnumbers,amsmath,amssymb]{revtex4}

\usepackage {amsmath,amssymb,epsfig,color}
\usepackage {graphicx}% Include figure files
\usepackage {dcolumn}% Align table columns on decimal point
\usepackage {bm}% bold math

\begin{document}

\title{Itinerant in-plane magnetic fluctuations and many-body correlations in Na$_x$CoO$_2$}

\author {M.M.~Korshunov $^{1,2}$}
 \email {maxim@mpipks-dresden.mpg.de}
\author {I.~Eremin $^{2,3}$}
\author {A.~Shorikov $^4$}
\author {V.I.~Anisimov $^4$}
\author {M.~Renner $^3$}
\author {W.~Brenig $^3$}
 \affiliation {$^1$ L.V. Kirensky Institute of Physics, Siberian Branch of Russian Academy of Sciences, 660036 Krasnoyarsk, Russia}
 \affiliation {$^2$ Max-Planck-Institut f\"{u}r Physik komplexer Systeme, D-01187 Dresden, Germany}
 \affiliation {$^3$ Institute f\"{u}r Mathematische und Theoretische Physik, TU Braunschweig, 38106 Braunschweig, Germany}
 \affiliation {$^4$ Institute of Metal Physics, Russian Academy of Sciences-Ural Division, 620041 Yekaterinburg GSP-170, Russia}

\date{\today}

\begin{abstract}
Based on the {\it ab-initio} band structure for Na$_x$CoO$_2$
we derive the single-electron energies and the effective
tight-binding description for the $t_{2g}$ bands using projection
procedure.
Due to the presence of the next-nearest-neighbor hoppings a local
minimum in the electronic dispersion close to the $\Gamma$ point of
the first Brillouin zone forms. Correspondingly, in addition to a
large Fermi surface an electron pocket close to the $\Gamma$ point
emerges at high doping concentrations. The latter yields the new
scattering channel resulting in a peak structure of the itinerant
magnetic susceptibility at small momenta. This indicates dominant
itinerant in-plane ferromagnetic fluctuations above certain critical
concentration $x_m$, in agreement with neutron scattering data.
Below $x_m$ the magnetic susceptibility shows a tendency towards the
antiferromagnetic fluctuations. We further analyze the many-body
effects on the electronic and magnetic excitations using various
approximations applicable for different $U/t$ ratio.
\end{abstract}

\pacs{74.70.-b; 71.10.-w; 75.40.Cx; 31.15.Ar}

\maketitle

\section{Introduction \label{section:intro}}

The recent discovery of the superconductivity in hydrated lamellar
cobaltate Na$_x$CoO$_2 \cdot y$H$_2$O \cite{kt2003} raised
tremendous interest in the nature and symmetry of the
superconductive pairing in these materials. The phase diagram of
this compound, with varying electron doping concentration $x$ and
water intercalation $y$, is rich and complicated; in addition to
superconductivity, it exhibits magnetic and charge orders, and some
other structural transitions \cite{it1997,yw2000,mlf2004,bcs2004}.
The parent compound, Na$_x$CoO$_2$,  is a quasi-two-dimensional system
with Co in CoO$_2$ layers forming a triangular lattice where the Co-Co
in-plane distance is two times smaller than the inter-plane one. Na
ions reside between the CoO$_2$ layers and donate additional $x$
electrons to the layer, lowering the Co valence from Co$^{4+}$
(3$d^5$ configuration) to Co$^{3+}$ (3$d^6$ configuration) upon
changing $x$ from 0 (CoO$_2$) to 1 (NaCoO$_2$). The
hole in the $d$-orbital occupies one of the $t_{2g} $ levels, which
are lower than $e_g$ levels by about 2 eV \cite{djs2000}. The
degeneracy of the $t_{2g}$ levels is partially lifted by the
trigonal crystal field distortion which splits the former into the
higher lying $a_{1g}$ singlet and the lower two $e'_g$ states.

First principles LDA (local density approximation) and LDA+U band
structure calculations predict Na$_x$CoO$_2$ to have a large
Fermi surface (FS) centered around the $\Gamma=(0,0,0)$ point
with mainly $a_{1g} $ character and six hole pockets near the
${\rm K}=(0,\frac{4\pi}{3},0)$ points of the hexagonal Brillouin
zone of mostly $e'_g$ character  for a wide range of $x$ \cite{djs2000,kwl2004}.
At the same time, recent surface sensitive Angle-Resolved 
Photo-Emission Spectroscopy (ARPES) experiments 
\cite{mzh2004, hby2004, hby2005,dq2006} reveals a doping dependent 
evolution of the Fermi surface, which shows no sign of hole pockets 
for a wide range of Na concentrations, i.e. ($0.3 \le x \le 0.8)$.
Instead, the Fermi surface is observed to be centered around
the $\Gamma $ point and to have mostly $a_{1g}$ character. 
Furthermore, a dispersion of the valence band is measured 
which is only half of that calculated within the LDA. This 
indicates the importance of the electronic correlations in 
Na$_x$CoO$_2$.

Shubnikov-de Haas effect measurements revealed two 
well-defined frequencies in Na$_{0.3}$CoO$_2$, suggesting 
either the existence of Na superstructures or the presence of 
the $e'_g$ pockets \cite{lb2006}. Last possibility was found 
to be incompatible with existing specific heat data.
Also, within the LDA scheme the Na disorder was shown to destroy 
the small $e'_g$ pockets in Na$_{2/3}$CoO$_2$ because of their 
tendency towards the localization \cite{djs2006}.

The hole pockets are absent in the  LSDA+U (Local Spin Density
Approximation + Hubbard U) calculations \cite{pz2004}. However, in
this approach the insulating gap is formed by a splitting of the local
single-electron states due to spin polarization, resulting in a
spin polarized Fermi surface with an area twice as larger as ARPES
observes.

The dynamical character of the strong electron
correlations has been taken into account within Dynamical Mean Field
Theory (DMFT) calculations \cite{hi2005} and, surprisingly has led
to an enhancement of the area of the small Fermi surface pockets,
in contrast to the experimental observations. At the same time, the use
of the strong-coupling Gutzwiller approximation within the
multiorbital Hubbard model with fitting parameters \cite{sz2005}
yields an absence of the hole pockets at the Fermi surface.
According to these findings, the bands crossing the Fermi surface
have $a_{1g}$ character.

Concerning the magnetic properties, LSDA predicts Na$_x$CoO$_2$ to have
a weak intra-plane itinerant ferromagnetic (FM) state for nearly all
Na concentrations, $0.3 \leq x \leq 0.7$ \cite{djs2003}. On
contrary, neutron scattering finds A-type antiferromagnetic
order at $T_m \approx 22$K with an inter(intra)-plane
exchange constant $J_{c(ab)} = 12(-6)$meV and with
ferromagnetic ordering within Co-layer {\it only} for $0.75 \leq x
\leq 0.9$ \cite{atb2004,spb2005,lmh2005}.

In this paper we derive an effective low-energy model describing
the bands crossing the Fermi level on the basis of the LDA band
structure calculations. Due to the FS topology, inferred from LDA
band structure, the magnetic susceptibility $\chi_0({\bf q},\omega=0)$
reveals two different regimes. For $x < 0.56$ the susceptibility
shows pronounced peaks at the antiferromagnetic (AFM)
wave vector ${\bf Q}_{AFM}=\left\{\left(\frac{2\pi}{3},\frac{2\pi}{\sqrt{3}}\right),
\left(\frac{4\pi}{3},0\right) \right\}$ resulting in a tendency
towards in-plane $120^\circ$ AFM order. For for $x > 0.56$ the
susceptibility is peaked at small momenta near ${\bf Q}_{FM}=\left(0,0\right)$.
This clearly demonstrates the tendency of the system towards
an itinerant in-plane FM state. We find that the formation of the
electron pocket around the $\Gamma$ point is crucial for the
in-plane FM ordering at high doping concentrations. We further
analyze the role of the many-body effects calculated within the
Fluctuation-Exchange (FLEX), Gutzwiller, and Hubbard-I
approximations.

The paper is organized as follows. In Section~\ref{section:tbmodel}
the LDA band structure and tight-binding model parameters derivation
are described. The doping dependent evolution of the magnetic
susceptibility within the tight-binding model is presented in
Section~\ref{section:chi0}. The role of strong electron correlations
is analyzed in Section~\ref{section:sec}. The last Section
summarizes our study.

\section{Tight-binding model \label{section:tbmodel}}

The band structure of Na$_{0.61}$CoO$_2$ was obtained within the LDA
\cite{Kohn65} in the framework of TB-LMTO-ASA (Tight Binding
approach to the Linear Muffin-Tin-Orbitals using Atomic Sphere
Approximation) \cite{Andersen84} computation scheme. 
This compound crystallizes at 12K in the hexagonal structure ($P6_3/mmc$ 
symmetry group) with a=2.83176\AA  and c=10.84312\AA \cite{jorgensen2003}.
A displacement of Na atoms from its ideal sites $2d$ $(1/3,2/3,3/4)$
on about 0.2\AA are observed in defected cobaltates for both room
and low temperatures. This is probably due to the repulsion of the
randomly distributed Na atoms, locally violating hexagonal
symmetry~\cite{jorgensen2003}. In this study Na atoms were shifted
back to the high symmetry $2d$ sites. 
Oxygen was situated in the high-symmetry $2d$-position $(1/2,2/3,3/4)$. 
The obtained Co-O distance is 1.9066\AA which agrees well with experimentally 
observed one 1.9072(4)\AA \cite{jorgensen2003}.
This unit cell was used for all doping concentrations. 
The effect of the doping was taken into
account within the virtual crystal approximation where each Co site
has six nearest neighbor virtual atoms with fractional number of
valence electrons $x$ and a core charge $10+x$ instead of randomly
located Na. Note, that all core states of the virtual atom are left
unchanged and corresponds to Na ones. We have chosen Co $4s, 4p, 3d$
states,  $2s, 2p, 3d $ states of O and $3s, 3p, 3d$ states of Na as
the valence states for the TB-LMTO-ASA computation scheme. The radii
of atomic spheres are 1.99 a.u. for Co, 1.61 a.u. for oxygen, and
2.68 a.u. for Na. Two classes of empty spheres (pseudo-atoms without
core states) were added in order to fill the unit cell volume.

\begin{figure}
\includegraphics[width=1\linewidth]{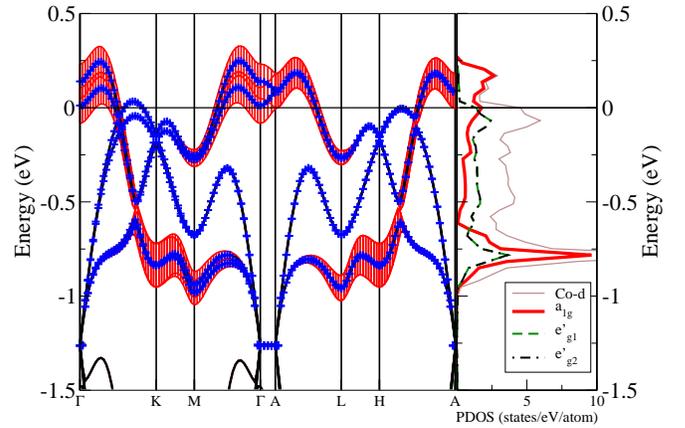}
\caption{(color online) Calculated near-Fermi level LDA band
structure and partial density of states (PDOS) for
Na$_{0.33}$CoO$_2$. The contribution of Co-$a_{1g}$ states is
denoted by the vertical broadening (in red) of the bands with
thickness proportional to the weight of the contribution. The
crosses indicate the dispersion of the bands obtained by projection
on the $t_{2g}$ orbitals. The horizontal line at zero energy denotes 
the Fermi level.}
\label{fig:bands}
\end {figure}
\begin{figure}
\includegraphics[width=1\linewidth]{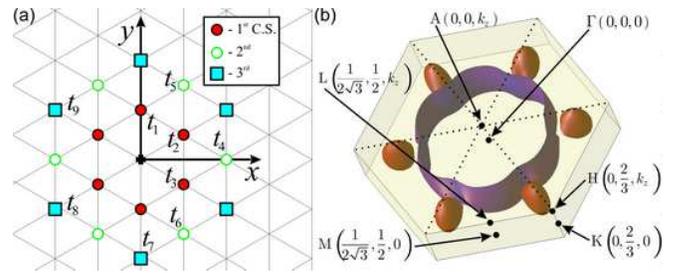}
\caption{(color online) (a) Schematic crystal structure of the
Co-layer in Na$_x$CoO$_2$ with hopping notations within the first
three coordination spheres (C.S.). (b) LDA-calculated Fermi surface
with cylindrical part (violet) having mostly $a_{1g}$ character and
six hole pockets (red) having mostly $e'_{g}$ character. $k_x$ and
$k_y$ coordinates of the symmetry points are given in units of $2\pi
/ a$ with $a$ being the in-plane lattice constant.}
\label{fig:hopp}
\end {figure}

In order to find an appropriate basis the occupation matrix was
diagonalized and its eigenfunctions were used as the new local
orbitals. This procedure takes into account the real distortion of
the crystal structure. The new orbitals are not pure trigonal
$a_{1g}$ and $e'_g$ orbitals but we still use the former notations
for the sake of simplicity.
288 $k$-points in the whole Brillouin zone were used for the band 
structure calculations (12x12x2 mesh for $k_x$, $k_y$, and $k_z$, correspondingly).

The bands crossing the Fermi level are shown in (Fig.~\ref{fig:bands}). 
One sees that they have mostly $a_{1g}$ character, consistent with 
previous LDA findings \cite{djs2000}.
Note, the small FS pockets near the K point with $e'_g$-symmetry
present at $x=0.33$ [see Fig.~\ref{fig:hopp}(b)] disappear for
higher doping concentrations because of the corresponding bands sink
below the Fermi level.
The difference in the dispersion along ${\rm K-M}$ and ${\rm L-H}$ directions is due to 
non-negligible interaction between CoO$_2$ planes. 
A small gap between Co-3$d$ and O-2$p$ states at about -1.25 eV present 
for x=0.61 disappears for x=0.33 due to the shift of the $d$-band to  
lower energy upon decreasing number of electrons.

In the following  we restrict ourselves to the model with the
in-plane hoppings inside CoO$_2$ layer to describe the doping
dependence of the itinerant in-plane magnetic order. Hence we
neglect bonding-antibonding (bilayer) splitting present in the
LDA-bands. This assumption seems to be justified  since the largest
interlayer hopping matrix element is an order of magnitude smaller
than the intra-plane one (0.012 eV vs. 0.12 eV).

\begin{table*}
\caption{Single-electron energies $\epsilon^{\alpha}$ (relative to
$\varepsilon^{a_{1g}}$) and in-plane hopping integrals
$t_{n}^{\alpha \beta}$ for Na$_{x}$CoO$_2$, where $x=0.33, 0.61,
0.7$. (all values are in eV)} \label{table:params}
\begin{ruledtabular}
\begin{tabular}{c|c|| c|c|c|c|c|c|c|c|c|c}
\multicolumn{2}{c||}{}& in-plane vector:
& (0, 1) & ($\frac{\sqrt{3}}{2}$, $\frac{1}{2}$) & ($\frac{\sqrt{3}}{2}$, -$\frac{1}{2}$) & ($\sqrt{3}$, 0)
& ($\frac{\sqrt{3}}{2}$, $\frac{3}{2}$) & ($\frac{\sqrt{3}}{2}$, -$\frac{3}{2}$) & (0, 2) & ($\sqrt{3}$, 1) & ($\sqrt{3}$, -1)      \\
\hline
$\alpha$ & $\epsilon^{\alpha}$& $\alpha \rightarrow \beta$
& $t_1^{\alpha \beta}$ & $t_2^{\alpha \beta}$ & $t_3^{\alpha \beta}$ & $t_4^{\alpha \beta}$
& $t_5^{\alpha \beta}$ & $t_6^{\alpha \beta}$ & $t_7^{\alpha \beta}$ & $t_8^{\alpha \beta}$ & $t_9^{\alpha \beta}$ \\
\hline \hline

\multicolumn{12}{c}{$x=0.33$}\\
\hline
$a_{1g}$  & 0.000& $a_{1g}  \rightarrow a_{1g}$  &  0.123& 0.123&  0.123&-0.022& -0.022&-0.021&-0.025 &-0.025& -0.025 \\
          &      & $a_{1g}  \rightarrow e'_{g1}$ & -0.044& 0.089& -0.044& 0.010&  0.010&-0.021& -0.021& 0.042&  -0.021 \\
$e'_{g1}$ &-0.053& $a_{1g}  \rightarrow e'_{g2}$ & -0.077& 0.000&  0.077& 0.018& -0.018& 0.000& -0.036& 0.000&  0.036 \\
          &      & $e'_{g1} \rightarrow e'_{g1}$ & -0.069&-0.005& -0.069& 0.018&  0.018&-0.026&-0.017 &-0.085 &-0.017 \\
$e'_{g2}$ &-0.053& $e'_{g1} \rightarrow e'_{g2}$ &  0.037& 0.000& -0.037&-0.026&  0.026& 0.000&-0.039 & 0.000 & 0.039 \\
          &      & $e'_{g2} \rightarrow e'_{g2}$ & -0.026&-0.090& -0.027&-0.011 & -0.011& 0.033&-0.062& 0.006 & -0.062 \\

\hline

\multicolumn{12}{c}{$x=0.61$}\\
\hline
$a_{1g}$  & 0.000& $a_{1g}  \rightarrow a_{1g}$  &  0.110& 0.110&  0.110&-0.019& -0.019&-0.019&-0.023&-0.023&-0.023\\
          &      & $a_{1g}  \rightarrow e'_{g1}$ & -0.050& 0.100& -0.050& 0.008&  0.008&-0.016&-0.017& 0.035&-0.017\\
$e'_{g1}$ &-0.028& $a_{1g}  \rightarrow e'_{g2}$ &  0.087& 0.000& -0.087&-0.014&  0.014& 0.000& 0.030&-0.000&-0.030\\
          &      & $e'_{g1} \rightarrow e'_{g1}$ & -0.069&-0.031& -0.069& 0.015&  0.015&-0.022&-0.016&-0.076&-0.016\\
$e'_{g2}$ &-0.028& $e'_{g1} \rightarrow e'_{g2}$ & -0.022& 0.000&  0.022& 0.021& -0.021& 0.000& 0.035& 0.000&-0.035\\
          &      & $e'_{g2} \rightarrow e'_{g2}$ & -0.044&-0.081& -0.044&-0.009& -0.009& 0.027&-0.056& 0.005&-0.056\\

\hline

\multicolumn{12}{c}{$x=0.7$}\\
\hline
$a_{1g}$  & 0.000& $a_{1g}  \rightarrow a_{1g}$  &  0.105& 0.105&  0.105&-0.018& -0.018&-0.018&-0.022&-0.022&-0.022\\
          &      & $a_{1g}  \rightarrow e'_{g1}$ & -0.052& 0.104& -0.052& 0.007&  0.007&-0.015&-0.016& 0.033&-0.016\\
$e'_{g1}$ &-0.019& $a_{1g}  \rightarrow e'_{g2}$ & -0.090& 0.000& -0.090& 0.013& -0.013& 0.000&-0.028& 0.000& 0.028\\
          &      & $e'_{g1} \rightarrow e'_{g1}$ & -0.068&-0.039& -0.068& 0.014&  0.014&-0.020&-0.015&-0.073&-0.015\\
$e'_{g2}$ &-0.019& $e'_{g1} \rightarrow e'_{g2}$ &  0.016& 0.000& -0.016&-0.020&  0.020& 0.000&-0.034& 0.000& 0.034\\
          &      & $e'_{g2} \rightarrow e'_{g2}$ & -0.048&-0.077& -0.049&-0.009& -0.009& 0.026&-0.054& 0.005&-0.054\\
\hline

\end{tabular}
\end{ruledtabular}
\end{table*}
To construct the effective Hamiltonian and to derive the effective
Co-Co hopping integrals $t_{fg}^{\alpha \beta}$ for the
$t_{2g}$-manifold we apply the projection procedure \cite{Marzari97,
Anisimov05}. Here, $(\alpha \beta)$ denotes a pair of orbitals,
$a_{1g}$, $e'_{g1}$ or $e'_{g2}$. The indices $f$ and $g$ correspond
to the Co-sites on the triangular lattice. The obtained hoppings and
the single-electron energies are given in Table~\ref{table:params}
for three difference doping concentrations. A comparison between the
bands obtained using projection procedure and the LDA bands is shown
in Fig.~\ref{fig:bands} confirming the Co-$t_{2g}$ nature of the
near-Fermi level bands \cite{djs2000, johannes2004}. For simplicity
we have enumerated site pairs with $n=0, 1, 2, ...$, $t_{fg}^{\alpha
\beta} \rightarrow t_n^{\alpha \beta}$ (see Fig.~\ref{fig:hopp}(a)
and the correspondence between in-plane vectors and index $n$ in
Table~\ref{table:params}). Due to the $C_3$ symmetry of the lattice,
the following equalities apply: $| t_3^{\alpha \beta} | = |
t_1^{\alpha \beta} |$, $| t_5^{\alpha \beta} | = | t_4^{\alpha
\beta} |$, $| t_9^{\alpha \beta} | = | t_7^{\alpha \beta} |$. In
addition $t_1^{\alpha \beta} = t_2^{\alpha \beta}$ for $a_{1g}
\rightarrow a_{1g}$ hoppings, which, however, does not hold for
$e'_{g1,2}$ orbitals. Since the hybridization between the $a_{1g}$
and the $e'_{g}$ bands is not small, a simplified description of the
bands crossing the Fermi level in terms of the $a_{1g}$ band only
(neglecting $e'_{g}$ band and the corresponding hybridizations, see
for example Ref.~\cite{kk2005}) may lead to an incorrect result due
to a higher symmetry of the $a_{1g}$-band.

In summary, the free electron Hamiltonian for CoO$_2$-plane in a hole
representation is given by:
\begin{equation}
H_0 = - \sum\limits_{{\bf k},\alpha ,\sigma } {\left( {\varepsilon
^\alpha - \mu } \right)n_{{\bf k} \alpha \sigma } } -
\sum\limits_{{\bf k}, \sigma} \sum\limits_{\alpha, \beta}
t_{{\bf k}}^{\alpha \beta } d_{{\bf k} \alpha \sigma }^\dag d_{{\bf k} \beta \sigma}.
\label{eq:H0}
\end{equation}
where $d_{{\bf k} \alpha \sigma}$ ($d_{{\bf k} \alpha \sigma}^\dag$)
is the annihilation (creation) operator for the hole with momentum
${\bf k}$, spin $\sigma$ and orbital index $\alpha$, $n_{{\bf k}
\alpha \sigma} = d_{{\bf k} \alpha \sigma}^\dag d_{{\bf k} \alpha
\sigma}$, and $t_{{\bf k}}^{\alpha \beta }$ is the Fourier transform
of the hopping matrix element, $\epsilon^{\alpha}$ is the single-electron 
energies, and $\mu$ is the chemical potential.
Introducing matrix notations, $\left( {\hat {t}_{{\bf k}} } \right)_{\alpha \beta} 
= t_{{\bf k}}^{\alpha \beta}$ and $\left( {\hat {t}_n } \right)_{\alpha \beta} =
t_n^{\alpha \beta}$, the hoppings matrix elements in the momentum
representation are given by:
\begin{eqnarray}
\hat{t}_{{\bf k}} &=& 2 \hat{t}_1 \cos k_2 + 2 \hat{t}_2 \cos k_3 + 2 \hat {t}_3 \cos k_1 \nonumber \\
&+& 2 \hat{t}_4 \cos (k_1 + k_3) + 2 \hat{t}_5 \cos (k_2 + k_1) + 2 \hat{t}_6 \cos (k_1 - k_2) \nonumber \\
&+& 2 \hat{t}_7 \cos 2 k_2 + 2 \hat{t}_8 \cos 2 k_3 + 2 \hat{t}_9 \cos 2 k_1,
\end{eqnarray}
where $k_1 = \frac{\sqrt 3}{2} k_x - \frac{1}{2} k_y$,  $k_2 = k_y$,
$k_3 = \frac{\sqrt 3}{2} k_x + \frac{1}{2} k_y$.

\begin{figure}
\includegraphics[width=1\linewidth]{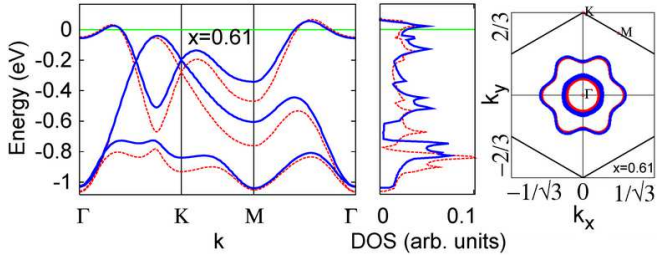}
\caption{(color online) Calculated tight-binding energy dispersion,
the density of states (DOS), and the Fermi surface for
Na$_{0.61}$CoO$_2$ within the rigid-band approximation with {\it ab
initio} parameters for $x=0.61$ (the solid blue curve) and for
$x=0.33$ (red dashed curve). The horizontal (green) line denotes the
chemical potential $\mu$ for $x=0.61$.} \label{fig:diffparams}
\end {figure}

Note, the parameters do not change significantly upon changing the
doping concentration.
In Fig.~\ref{fig:diffparams} we show two results of the 
rigid-band approximation with the Hamiltonian (\ref{eq:H0})
and the hopping values obtained in LDA calculation for two 
different doping concentrations, $x=0.33$ and $x=0.61$ (see Table~\ref{table:params}).
The doping concentration used to calculate the chemical potential 
$\mu$ was fixed to be $x=0.61$ for both Hamiltonians.
Although one finds the pronounced
differences in the dispersion around the ${\rm M}$-point, they are
small around the FS. Since most of the physical quantities are
determined by the states lying close to the Fermi level, we can
safely ignore the small differences of the band structure and
describe the doping evolution of the Na$_x$CoO$_2$ by simply varying
the chemical potential. In the following we will use {\it ab initio}
parameters calculated for $x=0.33$ and change the chemical potential
to achieve different doping concentrations.

Within the rigid band approximation the $e'_g$ hole pockets are well
below the Fermi level for $x \geq 0.41$. Most important however,
we find the local minimum of the band dispersion around the $\Gamma$
point (see Fig.~\ref{fig:diffparams}) to yield an inner FS
contour centered around this point. The area of this electron FS
pocket increases upon increasing the doping concentration $x$. As we
will show later, the main reason for the local minimum around the
$\Gamma$ point is the presence of the next-nearest-neighbor hopping
integrals which enter our tight-binding dispersion. Although this
minimum is not yet directly observed by ARPES experiments, note
that the inner FS contour would reduce the total FS volume and
therefore may explain why the volume of the FS observed in ARPES so
far is larger than it follows from Luttinger's theorem
\cite{geck2006}. Furthermore, an emergence of this pocket would
influence the Hall conductivity at high doping concentrations which
is interesting to check experimentally.

Note, the appearance of the inner contour of the FS around the
$\Gamma$ point for large doping concentrations is not unique to our
calculations, previously it has been obtained within the LDA
calculations for a single Co-layer per unit cell\cite{pz2004}.

\section{Magnetic susceptibility \label{section:chi0}}

\begin{figure}
\includegraphics[width=1\linewidth]{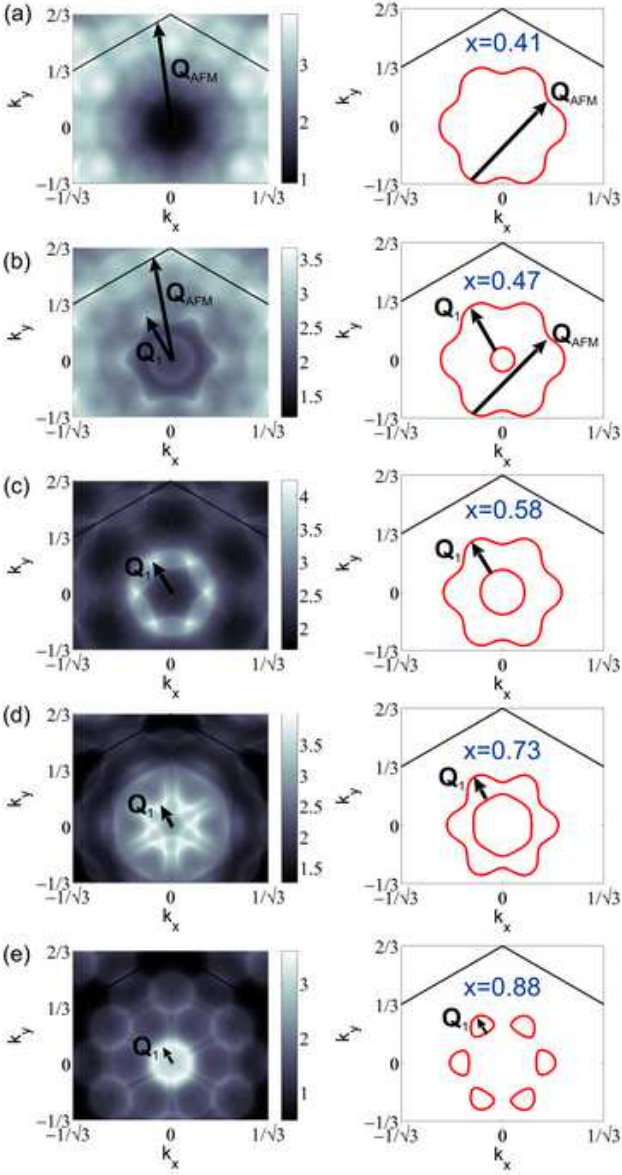}
\caption{(color online) The contour plot of the real part of the
magnetic susceptibility ${\rm Re} \chi_0 ({\bf k},\omega=0)$ as a
function of the momentum in units of $2\pi / a$ (left), and the
Fermi surface for corresponding doping concentration $x$ (right).
The arrows indicate the scattering wave vectors ${\bf Q_i}$ as
described in the text.} \label{fig:chi0}
\end {figure}

To analyze the possibility of the itinerant magnetism we calculate
the magnetic susceptibility $\chi_0({\bf q},\omega=0)$ based on the
Hamiltonian $H_0$. The doping-dependent evolution of the peaks in
${\rm Re} \chi_0({\bf q},0)$ is shown in Fig.~\ref{fig:chi0}. At
$x=0.41$ the $e'_g$ bands are below the Fermi level, and the FS has
the form of the rounded hexagon. It results in a number of nesting
wave vectors around the antiferromagnetic wave vector ${\bf
Q}_{AFM}$. The corresponding broad peaks in the ${\rm Re}
\chi_0({\bf q},0)$ appear around ${\bf Q}_{AFM}$, indicating the
tendency of the electronic system towards an $120^\circ$ AFM SDW
ordered state \cite{mdj2004}. Upon increasing doping, the Fermi
level crosses the local minimum at the $\Gamma$ point, resulting in
an almost circle inner FS contour. As soon as this change of the
FS topology occurs, the scattering at the momentum ${\bf Q}_{AFM}$
is strongly suppressed at $x_m \geq 0.56$. Simultaneously, a new
scattering vector, ${\bf Q}_1$, at small momenta appears.
Correspondingly, the magnetic susceptibility peaks at small momenta,
indicating the tendency of the magnetic system towards itinerant SDW
order with small momenta. The relevance of the local minimum around
the $\Gamma$ point for the formation of the scattering at small
momenta was originally found in Ref.~\cite{kk2005}.

For large $x$ the area of the inner FS contour increases leading to
a further decrease of the ${\bf Q}_1$. Observe that for $x \approx
0.88$, the FS topology again changes yielding six distant FS
contours that moves ${\bf Q}_1$ further to zero momenta. The
scattering at small momenta seen in the bare magnetic susceptibility
for $x > x_m$ is qualitatively consistent with the ferromagnetic
ordering at ${\bf Q}_{FM}=\left(0,0\right)$, observed in the neutron
scattering experiments \cite{atb2004,spb2005,lmh2005}.

\section{Effects of strong electron correlations \label{section:sec}}

It is important to understand the impact of electronic correlations
on the magnetic instabilities obtained within the rigid band
approximation. Since obtained magnetic susceptibility depends
mostly on the topology of the FS one expects that the behavior
shown in Fig.~\ref{fig:chi0} will be valid even if one
consider an RPA susceptibility with an interaction
term $H_{int}$ taken into account, at least in the case
if the only interaction is the on-site Hubbard repulsion $U$.
The only
difference would be a shift of the critical concentrations
$x_m$, at which the FS topology changes and tendency to the
AFM order changes towards the tendency to the FM ordered state.
Similar to Refs. \cite{sz2005, mi2005} we add the
on-site Coulomb interaction terms to Eq.(\ref{eq:H0}). At present,
it is not completely clear to which extent the electronic correlations
governs the low-energy properties in Na$_x$CoO$_2$ due to multi-orbital
effects in this compound which complicates the situation. Therefore,
in the following we discuss three different approximations valid for
different $U/t$ ratio.

\subsection{Hubbard-I approximation\label{section:hubbard}}

To analyze the regime of strong electron correlations we project
the doubly occupied states out and formulate an effective model
equivalent to the Hubbard model with an infinite value of $U$.
This approximation could be justified by the large ratio of
the on-site Coulomb interaction on the CoO$_2$ cluster
$U$ with respect to the bandwidth $W$. In the
atomic limit the local low-energy states on the Co sites are the
vacuum state $\left| 0 \right> $ and the single-occupied hole states
$\left| {a \sigma } \right> $, $\left| {e_1 \sigma } \right> $,
$\left| {e_2 \sigma } \right>$.
\begin{figure}
\includegraphics[width=0.8\linewidth]{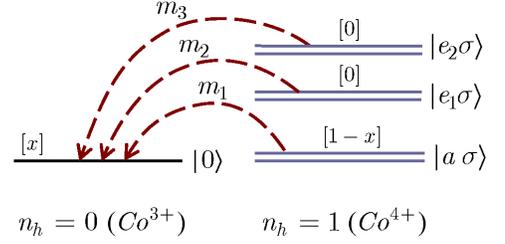}
\caption{(color online) A schematic picture of the local atomic
states on Co and the single-particle excitations in Na$_x$CoO$_2$. Here
$n_h$ stands for number of holes, $m_i$ enumerates single-particle
excitations. The filling factor of the corresponding state upon
changing the doping concentration $x$ is given in square brackets.}
\label{fig:localstates}
\end {figure}
The single-particle hole excitations  and local atomic states are
shown in Fig.~\ref{fig:localstates}. The simplest way to describe
the quasiparticle excitations between these states is to use the
projective Hubbard $X$-operators that take the
no-double occupancy constraint into account automatically \cite{Hubbard1964}:
$X_f^m \leftrightarrow X_f^{p,q} \equiv \left| p \right> \left< q
\right|$, where index $m \leftrightarrow (p,q)$ enumerates
quasiparticles. There is a simple correspondence between the
fermionic-like $X$-operators and single-electron
creation-annihilation operators: $d_{f\alpha \sigma } =
\sum\limits_m { \gamma_{\alpha \sigma}(m) X_f^m }$, where
$\gamma_{\alpha \sigma }(m)$ determines the partial weight of a
quasiparticle $m$ with spin $\sigma$ and orbital index $\alpha$. In
these notations the Hamiltonian of the Hubbard model in the limit $U
\to \infty$ has the form:
\begin{eqnarray}
H = - \sum\limits_{f,p} {\left( {\varepsilon^p - \mu } \right)X_f^{p,p} }
- \sum\limits_{f \ne g} {\sum\limits_{m,m'} {t_{fg}^{mm'} {X_f^{m}}^\dag X_g^{m'} } }.
\end{eqnarray}

\begin{figure}
\includegraphics[width=1\linewidth]{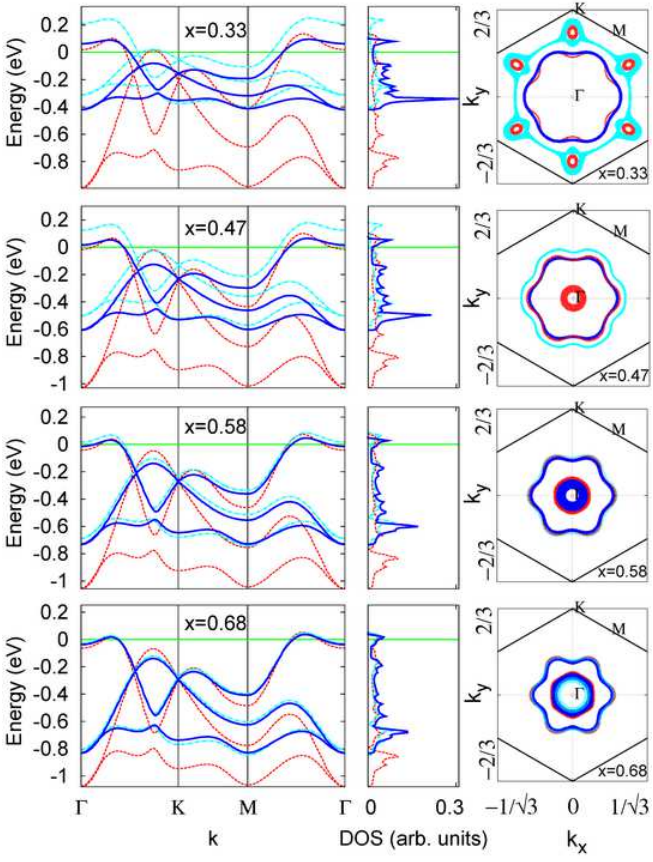}
\caption{(color online) Calculated band structure and the Fermi
surface topology for Na$_x$CoO$_2$ for $x = 0.33, 0.47, 0.58, 0.68$.
The dashed (red), solid (blue) and dash-dotted (cyan) curves
represent the results of the rigid-band, the Gutzwiller, and the
Hubbard-I approximations, respectively. The horizontal (green) line
denotes the position of the chemical potential $\mu$.}
\label{fig:approximations_bands}
\end {figure}
To study a quasiparticle energy spectrum of the system and its thermodynamics
we use the Fourier transform of the two-time retarded Green function in the
frequency representation, $ G_{\alpha \sigma}({\bf k},E) \equiv \left<\left< d_{{\bf k}
\alpha \sigma} \left| d_{{\bf k} \alpha \sigma}^\dag \right. \right>\right>_E$.
This can be rewritten as: $G_{\alpha \sigma}({\bf k},E) = \sum\limits_{m,m'} {\gamma_{\alpha \sigma}(m)
\gamma_{\beta \sigma}^*(m')} D^{mm'}({{\bf k}},E)$,
where
$D^{mm'}({\bf k},E) = \left<\left< X_{{\bf k}}^m \left| {X_{{\bf
k}}^{m'}}^\dag \right. \right>\right>_E$ is the matrix Green function in the $X$-operators representation.

Using the diagram technique for Hubbard X-operators
\cite{zaitsev1975,izumov1991} one obtains the generalized Dyson
equation \cite{ovchinnikov_book2004}:
\begin{equation}
\hat{D}({\bf k},E) = \left[ {\hat{G}_0^{-1}(E) - \hat{P}({\bf k},E) \hat{t}_{{\bf k}} + \hat{\Sigma}({\bf k},E)} \right]^{-1} \hat{P}({\bf k},E).
\label{eq:D}
\end{equation}
Here, $\hat{G}_0^{-1}(E)$ stands for the (exact) local Green
function, $G_0^{mm'}(E) = \delta_{mm'} / \left[ {E -
\left({\varepsilon_p - \varepsilon_q } \right)} \right]$. In the
Hubbard-I approximation the self-energy $\hat{\Sigma}({\bf k},E)$
is equal to zero and the strength operator $\hat{P}({\bf k},E)$ is
replaced by the sum of the occupation factors, $P^{mm'}({\bf k},E)
\to P^{mm'} = \delta_{mm'} \left[ \left< X_f^{p,p} \right> + \left<
X_f^{q,q} \right> \right]$, $m=m(p,q)$. 
Here '$\left< ... \right>$' stands for the usual thermodynamic average.
Thus, one obtains:
\begin{equation}
\hat{D}^{(0)}({\bf k},E) = \left[ {\hat{G}_0^{-1} - \hat{P} \hat{t}_{{\bf k}}} \right]^{-1} \hat{P}.
\label{eq:D0}
\end{equation}

In the paramagnetic phase the occupation factors are: 
$\left<X_f^{0,0} \right> = x$, 
$\left< X_f^{a \sigma,a \sigma} \right> = \frac{1 - x}{2}$, 
$\left< X_f^{e_{1,2} \sigma,e_{1,2} \sigma} \right> = 0$ 
which yields the diagonal form of the strength
operator, $\hat{P} = {\rm diag}\left( \frac{1 + x}{2}, x, x \right)$. 
Therefore, the quasiparticle bands formed by the $a_{1g} \to a_{1g}$ 
hoppings will be renormalized by the $(1 + x)/2$ factor,
while the quasiparticle bands formed by the $e'_g$ hopping elements
will be renormalized by $x$.

In Fig.~\ref{fig:approximations_bands} the quasiparticle spectrum,
the DOS, and the FS are displayed in different approximations.
Within Hubbard-I approximations one finds the narrowing of the bands
with lowering the doping concentration $x$ due to doping dependence
of the quasiparticle's spectral weight introduced by the strength
operator $\hat{P}$. However, the doping evolution of the FS is
qualitatively similar to that in the rigid-band picture. Namely, the
bandwidth reduction and the spectral weight renormalization do not
change the topology of the FS. As a result, the presence of the
strong electronic correlations within Hubbard-I approximation do not
change qualitatively our results for the bare susceptibility.
Quantitatively, the critical concentration $x_m$ shifts towards
higher values of the doping and becomes $x_m \approx 0.68$. The
reason for this shift is the band narrowing and the renormalization
of the quasiparticle's spectral weight, which enters the equation
that determines the position of the chemical potential $\mu$.

Luttinger's theorem, which holds for a perturbative expansion of the
Green's function in terms of the interaction strength is violated
within the Hubbard-I approximation.
This violation is due to the renormalization of the spectral weight
of the Green function by the occupation factors in the strength
operator in Eq.~(\ref{eq:D}). This is the reason why in spite of the
$e'_g$ band narrowing the $e'_g$ hole pockets at the Fermi surface
are still present at $x=0.33$.

\subsection{Gutzwiller approximation \label{section:gutzwiller}}

The Gutzwiller approximation
\cite{gutzwiller1963,gebhard1990,kotliar1986} for the Hubbard model
provides a good description for the correlated metallic system. Its
 multiband generalization was
formulated in Ref.~\cite{gebhard1997}. In this approach, the
Hamiltonian describing the interacting system far from the
metal-insulator transition for $U >> W$
\begin{equation}
H = H_0 + \sum\limits_{f, \alpha} U_\alpha n_{f \alpha \uparrow} n_{f \alpha \downarrow},
\label{eq:HG}
\end{equation}
with $H_0$ being the free electron Hamiltonian (\ref{eq:H0}),
is replaced by the effective non-interacting Hamiltonian:
\begin{eqnarray}
H_{eff} &=& - \sum\limits_{f, \alpha, \sigma}
\left( \varepsilon^{\alpha} + \delta\varepsilon^{\alpha\sigma} - \mu \right) n_{f \alpha \sigma} \nonumber\\
&-& \sum\limits_{f \ne g, \sigma} \sum\limits_{\alpha, \beta}
{\tilde t_{fg}}^{\alpha \beta} d_{f \alpha \sigma}^\dag  d_{g \beta \sigma} + C.
\label{eq:Heff}
\end{eqnarray}
Here, ${\tilde t_{fg}}^{\alpha \beta} = t_{fg}^{\alpha \beta}
\sqrt{q_{\alpha \sigma}} \sqrt{q_{\beta \sigma}}$ is the
renormalized hopping, $q_{\alpha \sigma} = \frac{x}{1 - n_{\alpha
\sigma}}$, $n_{\alpha \sigma} = \left< \Psi_0 \right| n_{f \alpha
\sigma} \left| \Psi_0 \right> \equiv \left< n_{f \alpha \sigma}
\right>_0$ is the orbital's filling factors, $x = 1 -
\sum\limits_{\alpha \sigma} n_{\alpha \sigma}$ is the equation
for the chemical potential.
$\delta\varepsilon^{\alpha\sigma}$ are the
Lagrange multipliers yielding the correlation induced
shifts of the single-electron energies. The constant $C$ is
determined from the condition that the ground state energy
is the same for both Hamiltonians
\begin{equation}
\left< \Psi_0 \right| H_{eff} \left| \Psi_0 \right> = \left< \Psi_g \right| H \left| \Psi_g \right>,
\label{eq:Psi0Psig}
\end{equation}
where $\left| \Psi_0 \right>$ is the wave function of the free electron
system (\ref{eq:Heff}), and $\left| \Psi_g \right>$ is the Gutzwiller
wave function for the Hamiltonian (\ref{eq:HG}).

The Lagrange multipliers are determined by minimizing the energy,
\begin{eqnarray}
\left< \Psi_0 \right| H_{eff} \left| \Psi_0 \right> &=&  
- \sum\limits_{\alpha, \sigma} \left( \varepsilon^{\alpha} + \delta\varepsilon^{\alpha \sigma}
- \mu \right) \left< n_{f \alpha \sigma} \right>_0
 \nonumber\\
&-& \sum\limits_{f \ne g, \sigma} \sum\limits_{\alpha, \beta} {\tilde t_{fg}}^{\alpha \beta}
\left< d_{f \alpha \sigma}^\dag d_{g \beta \sigma} \right>_0 + C,
\end{eqnarray}
with respect to the orbital filling factors $n_{\alpha \sigma}$.
Here $C = \sum\limits_{\alpha, \sigma} \delta\varepsilon^{\alpha \sigma} n_{\alpha \sigma}$,
as determined from Eq.~(\ref{eq:Psi0Psig}).
This results in the following expression for the single-electron energy renormalizations:
\begin{equation}
\delta\varepsilon^{\alpha \sigma} = \frac{1}{2 \left(1 - n_{\alpha \sigma} \right)}
\sum\limits_{f \ne g, \beta} {\tilde t}_{fg}^{\alpha \beta} \left< d_{f \alpha \sigma}^\dag d_{g \beta \sigma} \right>_0.
\label{eq:vareps}
\end{equation}
It is this energy shift that forces the $e'_g$ FS hole pockets to
sink below the Fermi energy \cite{sz2005}, which is clearly seen in
the doping-dependent evolution of the quasiparticle dispersion and
the FS as obtained within Gutzwiller approximation
(Fig.~\ref{fig:approximations_bands}). Although the narrowing of the
bands due to strong correlations is similar to the one found in the
Hubbard-I approximation, the FS obeys Luttinger's theorem. 
Note, in contrast to the Hubbard-I approximation the relative
positions of the $t_{2g}$-bands are also renormalized by
$\delta\varepsilon^{\alpha \sigma}$.

At the same time, for $x>0.4$ the topology of the FS in the
Gutzwiller approximation is qualitatively the same as in the
rigid-band picture. The also yields similar results for the bare
susceptibility's doping dependence discussed in
Section~\ref{section:chi0}. The only effect of the strong
correlations for $\chi_0$ is the observed shift of the critical
concentration towards higher values, $x_m \approx 0.6$. This is
due to combined effect of the bands narrowing and the doping
dependence of the $a_{1g}$ and $e'_g$ band's relative positions,
determined by the Eq.~(\ref{eq:vareps}).

Note, for $x<0.4$, due to different FS topology that occurs in the
Gutzwiller approximation, the bare susceptibility differs from that
obtained in Ref.~\cite{mdj2004} where the strong renormalization of
the electronic bands removing $e'_g$ pockets away from the FS was
neglected.

\subsection{FLEX approximation  \label{section:flex}}

A certain disadvantage of the Gutzwiller and Hubbard-I like
approximations is that the dynamic character of electronic
correlations is not taken into account within these approaches. At
the same time, the momentum and frequency dependencies of the
self-energy $\Sigma ({\bf k}, \omega)$ play a crucial role, in
particular, for determining the low-energies excitations close to
the Fermi level. In this subsection we focus on the $a_{1g}$-band
with nearest and next-nearest hopping integrals only and
employ the single-band Fluctuation Exchange approximation (FLEX) \cite{flex}
which sums all
particle-hole(particle) ladder graphs for the generating functional
self-consistently valid for the intermediate strength of the
correlations. The FLEX equations for the single-particle Green
function $G$, the self-energy $\Sigma$, the effective interaction
$V$, the bare ($\chi^0$) and renormalized spin ($\chi^s$) and charge
($\chi^c$) susceptibilities read
\begin{eqnarray}
G_{\bf k}(\omega_n) &=& \left[ \omega_n - \tau_{\bf k} + \mu - \Sigma_{\bf k}(\omega_n) \right]^{-1}, \\
\Sigma_{\bf k}(\omega_n) &=& \frac{T}{N} \sum\limits_{{\bf p}, m} V_{\bf k-p}(\omega_n - \omega_m) G_{\bf p}(\omega_m), \\
V_{\bf q}(\nu_m) &=& U^2 \left[ \frac{3}{2} \chi^s_{\bf q}(\nu_m) +
\frac{1}{2} \chi^c_{\bf q}(\nu_m)
- \chi^0_{\bf q}(\nu_m) \right], \\
\chi^0_{\bf q}(\nu_m) &=& - \frac{T}{N} \sum\limits_{{\bf k}, n} G_{\bf k+q}(\omega_n + \nu_m) G_{\bf k}(\omega_n), \\
\chi^{s,c}_{\bf q}(\nu_m) &=& \frac{\chi^0_{\bf q}(\nu_m)}{1 \mp U \chi^0_{\bf q}(\nu_m)},
\end{eqnarray}
where $\omega_n = i \pi T (2n+1)$ and $\nu_m = i \pi T (2m)$. 
In the last equation the '$-$' sign in the denominator corresponds 
to the $\chi^{s}_{\bf q}(\nu_m)$, while the '$+$' sign corresponds 
to the $\chi^{c}_{\bf q}(\nu_m)$. 
We compute the Matsubara summations using 'almost real contour'
technique of Ref.~\cite{schmalian1996}. I.e., the contour integrals
are performed with a finite shift $i \gamma$ ($ 0 < \gamma < i T
\pi/2$) into the upper half-plane. All final results are
analytically continued from $\omega + i \gamma$ onto the real axis
$\omega + i 0^{+}$ by Pad\'{e} approximation. The following results
are based on FLEX solutions using a lattice of 64x64 sites with 4096
equidistant $\omega$-points in the energy range of $[-30,30]$. The
temperature has been kept at $T = 0.05 \tau$, where $\tau$ is
the hopping amplitude to the nearest neighbors for the $a_{1g}$ band
corresponding to
$t^{a_{1g} a_{1g}}_1 = t^{a_{1g} a_{1g}}_2 = t^{a_{1g} a_{1g}}_3$.
The Hubbard repulsion was set to $U = 8 \tau$.

\begin{figure}
\includegraphics[width=0.7\linewidth]{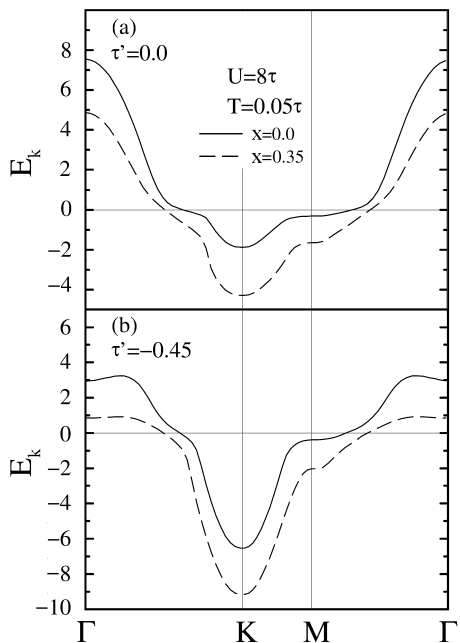}
\caption{Quasiparticle dispersion $E_{\bf k}$ (in units of $\tau$, relative to $\mu$)
within FLEX approximation for (a) $\tau'=0$ and (b) $\tau'=-0.45$
and for two doping concentrations.}
\label{fig:flexEk}
\end {figure}
\begin{figure}
\includegraphics[width=0.7\linewidth]{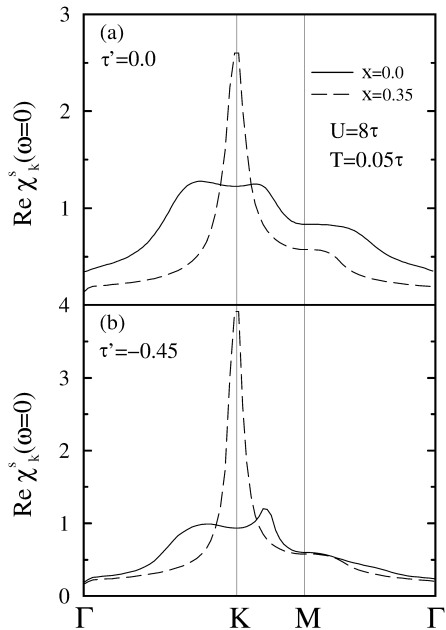}
\caption{Doping dependence of the static spin structure factor
${\rm Re}\chi_{{\bf k}}^s(\omega=0)$ for (a) $\tau'=0$ and (b) $\tau'=-0.45$.
Notice that for large $U=8\tau$ the commensurate peak at ${\rm K}$ point
is absent at a very low $x$.}
\label{fig:flexchi}
\end {figure}
\begin{figure}
\includegraphics[width=0.7\linewidth]{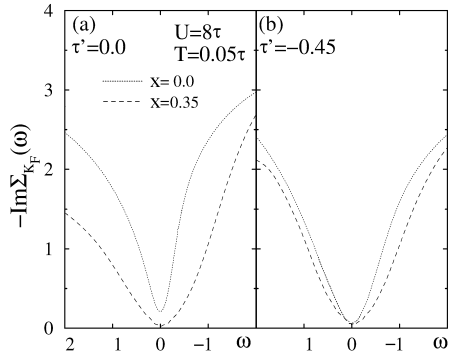}
\caption{Frequency dependence of quasiparticle self-energy $-{\rm Im}\Sigma_{{\bf k}}$
near the FS in direction $\Gamma - {\rm K}$ for (a) $\tau'=0$ and (b) $\tau'=-0.45$.}
\label{fig:flexSigma}
\end {figure}

Previously, the FLEX approximation has been applied successfully to the
study of superconductivity as well as spin and charge
excitations in Na$_x$CoO$_2$ \cite{kuroki,ogata}.
Complementary, we will focus on the quasiparticle dispersion and study
the impact of the momentum and frequency dependencies of the $\Sigma({\bf k}, \omega)$,
and the role played by the next-nearest hopping
integral, $\tau'$, corresponding to
$t^{a_{1g} a_{1g}}_4 = t^{a_{1g} a_{1g}}_5 = t^{a_{1g} a_{1g}}_6$.
The quasiparticle dispersion $E_{\bf k}$, which is determined from equation
$E_{\bf k} - \tau_{\bf k} + \mu - \Sigma_{\bf k}(E_{\bf k}) = 0$, is
shown in Fig.~\ref{fig:flexEk} for $\tau'=0$ and $\tau'=-0.45$, in
units of $\tau$. First, observe that the local minimum around the
$\Gamma$ point appears only if the next-nearest-neighbor hopping
$\tau'$ is included which agrees with our previous findings. In
addition, we obtain a pronounced mass enhancement of the order
of unity at the FS crossings - the so-called kink structure. This
enhancement is due to low-energy spin fluctuations which are present
in $\chi^s_{\bf q}(\omega)$ \cite{kuroki}.

To shed more light onto the two-dimensional spin correlations, in
Fig.~\ref{fig:flexchi} we display the static spin
structure factor ${\rm Re}\chi^{s}_{\bf k}(\omega=0)$
from the FLEX
for two different doping concentrations. As doping increases from $x=0$ towards
$x=0.35$ the maximum in the spin susceptibility $\chi^{s}_{\bf k}(\omega=0)$
moves towards the ${\rm K}$-point of the first BZ and develops into a
sharp and commensurate peak at ${\bf Q}_{AFM}$ and the incommensurate
spin fluctuations are suppressed. One may also note that the commensurate
peak is $\sim 60 \%$ larger for $\tau'=-0.45$ than for $\tau'=0$.
These results are consistent with those obtained in a previous sections.
We further notice smooth evolution of the quasiparticle dynamics with doping
in Na$_x$CoO$_2$ showing no sign of unusual behavior at $x=0$.

The frequency dependence of the imaginary part of the quasiparticle
self-energy, i.e. ${\rm Im}\Sigma_{{\bf k}}$, near the FS is shown in
Fig.~\ref{fig:flexSigma}. We find the self-energy to be nearly isotropic
along the FS with only a weak maximum occurring into the direction
of the commensurate spin fluctuations. Near the Fermi energy the
self-energy is clearly proportional to $\omega^2$ at low energies
for all dopings shown, which is indicative of the normal
Fermi-liquid behavior. This is in sharp contrast to the FLEX
analysis of the Hubbard model on the square lattice close to
half-filling. There one typically finds 'marginal' Fermi-liquid
behavior with ${\rm Im}\Sigma_{{\bf k}} \approx \omega$ over a wide
range of frequencies \cite{abk2000,wermbter1989}. Therefore, along
this line one is tempted to conclude that the normal state of the
superconducting cobaltates is more of conventional metallic nature
than in the High-T$_c$ cuprates. This is even more so, if one
realizes from Fig.~\ref{fig:flexSigma} that the quasiparticle
scattering rate displays its smallest curvature for $x=0.35$, which
implies the quasiparticles to be rather well defined there. For
lower $x$ proximity of the FS to the van Hove singularity (see flat
region of dispersion in Fig.~\ref{fig:flexEk}) enhances both the
absolute value of ${\rm Im}\Sigma_{{\bf k}} \propto \omega$ as well
as the curvature. This effect is most pronounced for $\tau'=0$.

\section{Conclusion \label{section:conclusion}}

To conclude, we have calculated the doping dependent magnetic
susceptibility in the tight-binding model with {\it ab-initio} calculated
parameters. We find, that at a critical doping concentration, $x_m$,
electron pocket develops on the FS in the center of the Brillouin zone.
For $x < x_m$, the system shows a tendency towards an $120^\circ$ AFM
ordered state, while for $x > x_m$ a peak in the magnetic susceptibility
forms at small wave vectors indicating a strong tendency towards
an itinerant FS state. Within a tight-binding model we have estimated
$x_m$ to be approximately $0.56$.
Analyzing the influence of strong Coulomb repulsion and the corresponding
reduction of the bandwidth and the quasiparticle spectral weight in
the strong-coupling Hubbard-I and Gutzwiller approximations, we
have shown that the critical concentration changes to $x_m \approx 0.68$
and $x_m \approx 0.6$, respectively. At the same time, the
underlying physics of the formation of the itinerant FM state
remains the same.

We neglected the bonding-antibonding splitting due to the
3-dimensionality in the non-intercalated compounds. This splitting
was taken into account in Ref.~\cite{kuroki}, where within the FLEX
approximation the single $a_{1g}$-band Hubbard model was considered.
The results obtained also suggest a tendency to FM
fluctuations for high doping concentrations. The presence of a
local band minimum around the $\Gamma$ point played a crucial role,
similar to our present study.

To analyze the low-energy quasiparticle properties at low doping
concentrations we have employed the single-band Hubbard model within the
FLEX approximation. We have found a significant FS mass enhancement of
order unity due to quasiparticle scattering from spin
fluctuations.  In contrast to the Hubbard model on the square
lattice we have found the quasiparticle scattering rate to display a
conventional Fermi-liquid type of energy dependence. We have also
shown that the static spin structure factor exhibits a large
commensurate peak at wave vector ${\bf Q}_{AFM}$ for doping
concentrations of $x\approx 0.35$. This response was found to be
significantly enhanced by the next-nearest-neighbor hopping,
emphasizing its significance.

\begin{acknowledgments}
We would like to thank G. Bouzerar, P. Fulde, S.G. Ovchinnikov, N.B. Perkins,
D. Singh, Ziqiang Wang, and V. Yushankhai for useful discussions,
I. Mazin for critical reading of the manuscript,
and S. Borisenko for sharing with us the experimental
results prior to publication. M.M.K. acknowledge support form
INTAS (YS Grant 05-109-4891) and RFBR (Grants 06-02-16100, 06-02-90537-BNTS).
A.S. and V.I.A. acknowledge the financial support from RFBR (Grants
04-02-16096, 06-02-81017), and NWO (Grant 047.016.005).
\end{acknowledgments}

\end{document}